%% file: main.tex
\documentclass[journal]{IEEEtran}
\usepackage[linesnumbered,ruled,vlined]{algorithm2e} 
\usepackage{amsmath}
\usepackage{color}
\usepackage{graphicx}
\usepackage{cite}
\usepackage{subcaption}
\usepackage{booktabs}
\usepackage{lmodern}
\usepackage[T1]{fontenc}
\usepackage{multirow}
\usepackage{array}
\usepackage{mathptmx}  
\usepackage{newtxtext}
\usepackage{newtxmath}

\begin{document}

\title{QuanCrypt-FL: Quantized Homomorphic Encryption with Pruning for Secure  Federated Learning}

\author{Md Jueal Mia 
  and 
        M. Hadi Amini%\textsuperscript{orcid}
        , \textit{Senior~Member,~IEEE}% <-this % stops a space
\thanks{Md Jueal Mia and M. Hadi Amini are with The Knight Foundation School of Computing and Information Sciences, Florida International University, Miami, FL 33199. They are also with the Sustainability, Optimization, and Learning for InterDependent networks laboratory (solid lab). (Corresponding author:M. Hadi Amini, email: amini@cs.fiu.edu, hadi.amini@ieee.org)}
\thanks{This work was supported by the National Center for Transportation Cybersecurity and Resiliency (TraCR), a U.S. Department of Transportation National University Transportation Center, headquartered at Clemson University, Clemson, South Carolina, USA.}
\thanks{Manuscript received XXXX; revised XXXX.}} % <-this 

\markboth{}%
{Mia and Amini: QuanCrypt-FL: Quantized Homomorphic
Encryption with Pruning for Secure 
Federated Learning}

\maketitle

\begin{abstract}
Federated Learning has emerged as a leading approach for decentralized machine learning, enabling multiple clients to collaboratively train a shared model without exchanging private data. While FL enhances data privacy, it remains vulnerable to inference attacks, such as gradient inversion and membership inference, during both training and inference phases. Homomorphic Encryption provides a promising solution by encrypting model updates to protect against such attacks, but it introduces substantial communication overhead, slowing down training and increasing computational costs. To address these challenges, we propose QuanCrypt-FL, a novel algorithm that combines low-bit quantization and pruning techniques to enhance protection against attacks while significantly reducing computational costs during training. Further, we propose and implement mean-based clipping to mitigate quantization overflow or errors. By integrating these methods, QuanCrypt-FL creates a communication-efficient FL framework that ensures privacy protection with minimal impact on model accuracy, thereby improving both computational efficiency and attack resilience. We validate our approach on MNIST, CIFAR-10, and CIFAR-100 datasets, demonstrating superior performance compared to state-of-the-art methods. QuanCrypt-FL consistently outperforms existing method and matches Vanilla-FL in terms of accuracy across varying client. Further, QuanCrypt-FL achieves up to 9x faster encryption, 16x faster decryption, and 1.5x faster inference compared to BatchCrypt, with training time reduced by up to 3x.

\end{abstract}


\IEEEpeerreviewmaketitle

\input{Introduction}

\input{Related_Work}

\input{Threa_model}

\input{Methodology}

\input{Experiments_and_result_analysis}
\input{Discussion}

\section{Conclusion}
In this paper, we propose QuanCrypt-FL, a novel FL method that employs low-bit quantization and pruning with layer-wise model encryption to enhance both security and efficiency in FL. QuanCrypt-FL effectively mitigates inference attacks, such as gradient inversion and membership inference, by encrypting model updates and reducing communication overhead through quantization and pruning. Although HE is computationally expensive, our evaluation demonstrates that QuanCrypt-FL significantly improves computational performance and privacy protection compared to existing methods, achieving faster encryption, decryption, inference, and training times while maintaining accuracy levels comparable to those of vanilla FL method. This approach is scalable for cross-silo and cross-device applications where both security and utility are essential. It offers a communication-efficient and privacy-preserving solution suitable for real-world applications. Future work could explore adaptive structured pruning to further enhance scalability and efficiency in larger FL deployments.

\section*{Acknowledgement}
This work is based upon the work supported by the National Center for Transportation Cybersecurity and Resiliency (TraCR) (a U.S. Department of Transportation National University Transportation Center) headquartered at Clemson University, Clemson, South Carolina, USA. Any opinions, findings, conclusions, and recommendations expressed in this material are those of the author(s) and do not necessarily reflect the views of TraCR, and the U.S. Government assumes no liability for the contents or use thereof.

\bibliographystyle{IEEEtran}
\bibliography{References}

\end{document}

%% file: Introduction.tex
\section{Introduction} \label{sec:Introduction}

\IEEEPARstart Data is essential for research, driving innovation and discoveries. However, privacy laws like the General Data Protection Regulation (GDPR) and the California Consumer Privacy Act (CCPA) impose strict regulations that complicate how data is collected, used, and shared \cite{r1_baik2020data} \cite{r2_boyne2018data}. Although these laws provide exceptions for research, compliance with requirements such as consent, anonymization, and data minimization is mandatory, thereby making large-scale data collection both challenging and time-consuming \cite{r3_hintze2018science}. This creates significant hurdles for deep learning model training, which relies heavily on large datasets, as centralized data collection is often restricted by these regulations. As a result, researchers face difficulties navigating legal frameworks, which can substantially slow down the research process \cite{r3_hintze2018science}.

Federated Learning (FL) \cite{r4_mcmahan2017communication} offers a promising solution by enabling edge users or organizations to collaboratively train global models by sharing local parameters or gradients without sharing raw data. FL is categorized into cross-device and cross-silo settings. Cross-device FL involves many mobile or IoT devices with limited resources, while cross-silo FL includes fewer organizations with reliable communications and robust computing power \cite{r42_huang2022cross}. During FL training, clients share gradients with an honest server to protect data. {\color{black} However, numerous studies have highlighted challenges of  indirect data theft through membership inference attacks (MIA) \cite{r5_zhang2020gan, r6_gu2022cs} and gradient inversion attacks (GIA) \cite{r7_zhu2019deep, r8_zhao2020idlg, r9_geiping2020inverting, r10_liang2023egia, r11_li2022e2egi, r12_xu2022agic, r13_jeon2021gradient}.
These attacks demonstrate that even without direct access to raw data, an honest-but-curious (or semi-honest) \cite{r14_goldreich2001foundations} server, while following the protocol correctly during FL training, can still infer sensitive information by analyzing the exchanged gradients from the communication channel. Such attacks or vulnerabilities raise questions regarding the real-time applicability of FL in cross-device and cross-silo applications.}

Homomorphic Encryption (HE) \cite{r18_jin2023fedml, r19_hu2024maskcrypt, r20_hosseini2021secure, r21_gong2023multi} is a widely adopted technique in FL to address security vulnerabilities. HE schemes are categorized based on the number of operations allowed on encrypted data: Partial Homomorphic Encryption (PHE) \cite{r25_elgamal1985public, r26_paillier1999public}, Somewhat Homomorphic Encryption (SWHE) \cite{r27_boneh2005evaluating, r28_gentry2009fully}, and Fully Homomorphic Encryption (FHE) \cite{r28_gentry2009fully, r30_brakerski2011fully}. These schemes significantly enhance security during the upload and download of model parameters between the server and edge users or organizations in the training phases \cite{r31_ma2022privacy, r32_fang2021privacy}. HE has therefore been increasingly adopted in various sectors, including healthcare \cite{r34_wang2023ppflhe}, finance \cite{r35_ou2020homomorphic}, and autonomous systems \cite{r36_jia2021blockchain}. During training, edge users or organizations upload encrypted gradients or parameters using a public key, allowing the server to perform aggregation on the encrypted data without decryption. The updated global model is then sent back to the users for the next round, ensuring that gradients remain secure from the server. In cross-silo FL settings, the Paillier cryptosystem \cite{r26_paillier1999public} is particularly well-suited, as it provides strong privacy guarantees without compromising learning accuracy \cite{r38_zhang2020batchcrypt}, making it a popular choice for secure FL deployments \cite{r43_liu2019secure}. However, Paillier is vulnerable to quantum attacks, as its security relies on the hardness of integer factorization, which can be efficiently solved using Shor’s algorithm in polynomial time \cite{r44_shor1999polynomial}. To counter this, HE schemes like CKKS \cite{r47_cheon2017homomorphic}, BGV \cite{r45_brakerski2014leveled}, and BFV \cite{r46_fan2012somewhat}, which are based on the hardness of (Ring) Learning with Errors (LWE) problems, are quantum-resistant. Among these, CKKS is particularly advantageous for privacy-preserving machine learning, as it supports real number arithmetic and Single Instruction Multiple Data (SIMD) operations.

However, HE in FL incurs high computational costs due to the mathematical operations, such as addition and multiplication, involved in the training phase \cite{r37_jiang2021flashe}. To address the challenges of high communication costs of HE in FL, techniques like BatchCrypt \cite{r38_zhang2020batchcrypt} have been developed, utilizing quantization methods to reduce the size of data transmitted during training. By converting high-precision gradients into low-precision integers, these approaches efficiently reduce the communication overhead while maintaining model accuracy. However, BatchCrypt's use of additive encryption limits its capability to perform multiplicative operations on encrypted data, rendering it unsuitable for more advanced optimization techniques that rely on such operations. Additionally, despite its improvements, BatchCrypt remains computationally expensive for large models and datasets, and its design focus on cross-silo FL makes it less suitable for cross-device scenarios where there are many resource-constrained devices with unreliable connections. Further, Yang et al. \cite{r63_yang2024privacy} utilized the BatchCrypt encryption technique with sparsification to improve FL efficiency. By applying compressive sensing to non-dense layers, they reduced both computational and communication overhead in HE-based systems. Compressive sensing works by transforming model parameters into a sparse signal, where only the essential parameters are retained, and the less important ones are set to zero, enabling significant compression. However, this method excludes dense layers, which typically contain more personalized information, from compression to preserve model accuracy. While this selective compression reduces overhead, it limits the overall efficiency gains, as the large compression matrices for non-dense layers significantly increase memory consumption on resource-limited clients, making it challenging to balance the need for efficient processing with constrained resources.

Moreover, numerical overflow during quantization can deviate model performance from convergence, as improperly scaled or shifted values may lead to inaccurate gradient updates and hinder effective model training. Gradient pruning is another approach to reduce the computational complexity during training in FL. This technique eliminates communication redundancy and minimizes storage and inference time, thus enhancing efficiency in distributed systems \cite{r50_long2023feddip}.

In this paper, we propose a novel method called QuanCrypt-FL, designed for application development in distributed machine learning (e.g., FL). Our study is divided into three phases. First, we integrate a robust FHE scheme (e.g., CKKS) to mitigate vulnerabilities arising from attacks, such as MIA and GI attacks, during both the training and inference phases. In contrast to BatchCrypt, our approach involves encrypting the full model update layer-wise from the local client device, which significantly reduces the training time compared to BatchCrypt. Second, to address the high computational complexity of HE in FL, we implement a low-bit quantization technique to reduce upload costs for users or organizations, although this introduces challenges related to numerical overflows. To address these challenges, we propose a mean-based dynamic layerwise clipping technique. Finally, we incorporate unstructured pruning to deactivate less important neurons during the upload of local models and the download of global models, thereby reducing the number of floating-point operations (FLOPs). Together, these techniques aim to minimize storage requirements and inference time. Additionally, quantization and pruning help reduce security vulnerabilities, even if a client is compromised by attackers, as the sparsified model limits their ability to infer sensitive information from the global model. Gradient pruning, introduced by Zhu \textit{et al.} \cite{r7_zhu2019deep}, reduces small-magnitude gradients to zero, significantly mitigating the impact of Gradient Inversion Attacks (GIA). Pruning over 70\% of the gradients renders recovered images unrecognizable \cite{r7_zhu2019deep, r77_huang2021evaluating}, providing an additional layer of protection even in scenarios where edge clients are compromised. However, existing techniques have not fully addressed the vulnerabilities arising from the compromise of client models during HE implementation. As part of this, we conducted a well-known GIA attack on our decrypted global model to demonstrate the robustness of our method. To the best of our knowledge, this is the first time quantization and pruning are being implemented alongside FHE in FL. We successfully resolved inconsistencies related to the quantization process in our mechanism while maintaining the performance of the global model. By implementing a novel clipping technique to limit extreme values in the model updates before applying quantization, we ensured more stable and accurate results, leading to improved robustness of the model. The main contributions of this paper are provided below:

\begin{itemize}

\item We propose the efficient \textbf{QuanCrypt-FL mechanism}, which ensures robust privacy protection and resilience against inference attacks while minimizing computational complexity and reducing the  impact on global model performance. We apply FHE layer-wise to client model updates, securing them against inference attacks. Further, we employ quantization and pruning to reduce communication and computation overhead. Specifically, pruning enhances efficiency while simultaneously providing resilience against adversarial threats, such as Gradient Inversion Attacks (GIA). Even with a decrypted model from clients or servers, an attacker will not be able to infer information by analyzing gradient. We provide practical defense guarantees by simulating a GIA on the final global model to demonstrate the effectiveness of our method in defending against such attacks.

\item We integrate low-bit \textbf{quantization} and dynamic \textbf{pruning} with \textbf{HE} to enhance both efficiency and privacy. Quantization reduces the precision of model weights, resulting in a 3X reduction in storage usage. Pruning eliminates less important weights, further reducing aggregation costs and memory overhead. This combination optimizes the training process, achieving a 2X reduction in inference time compared to Vanilla FL and a 1.5X reduction compared to BatchCrypt, by minimizing parameters and computational complexity.

\item We propose a layerwise dynamic \textbf{mean-based clipping mechanism} to address numerical inconsistencies during the quantization process. This technique clips each layer's parameters based on their individual mean values, ensuring more precise handling of different weight distributions and improving stability and accuracy in the quantized model.

\item We conduct \textbf{extensive empirical simulations} on the MNIST, CIFAR-10, and CIFAR-100 datasets, demonstrating significant improvements in model accuracy, storage efficiency, encryption time, decryption time, inference time, and training time compared to \textbf{BatchCrypt} \cite{r38_zhang2020batchcrypt} a privacy-preserving mechanisms.

\end{itemize}

The remainder of this paper is structured as follows: Section II outlines the current state-of-the-art studies in related works. Section III introduces the overview of threat model with mathematical analysis. Section IV outlines the methodology, which includes the QuanCrypt-FL algorithm. Section V presents the experiments and analyzes the results, followed by Section VI, which presents a detailed discussion and assessment. Finally, Section VII provides the conclusion and future works of the paper.

%% file: Related_Work.tex
\section{ Background and Related Work} \label{sec:Related Work}

In this section, we highlight the privacy requirements in FL and survey existing techniques, including Differential Privacy (DP), Secure Multiparty Computation (SMC), and HE. We explore the challenges of applying machine learning to homomorphically encrypted models and how efficiently HE enhances security against an honest-but-curious server and external adversaries. 

The concept of DP \cite{r51_dwork2006our, r52_dwork2006calibrating} has emerged as a crucial framework for providing quantifiable guarantees against data leakage. In FL, DP can be classified into two categories: Central Differential Privacy (CDP) and Local Differential Privacy (LDP). CDP ensures that the aggregate model does not reveal a client's participation or any information about individual training samples \cite{r53_dwork2014algorithmic, r54_abadi2016deep, r55_mcmahan2017learning}. LDP overcomes CDP limitations by ensuring privacy without trusting the data curator. In LDP, clients individually perturb or encode their data before submitting it to the central server \cite{r56_kasiviswanathan2011can, r57_evfimievski2003limiting, r58_sun2020ldp}. This approach helps obscure sensitive information by introducing noise to the model parameters. However, the primary drawback of DP is that this noise can degrade the performance of the global model \cite{r22_naseri2020local}. The trade-off between utility and noise is a key challenge of DP, making it impractical for real-time applications. This is especially problematic in scenarios where high precision and responsiveness are critical, such as in healthcare or autonomous driving applications, where even small accuracy losses caused by noise can lead to dangerous outcomes.

SMC is an effective method to protect model parameters from potential attacks in FL. It allows participants to contribute their data for computation without disclosing it to others \cite{r60_zhao2019secure}. SMC has been implemented in FL to securely aggregate model parameters using privacy-preserving protocols \cite{r39_liu2020secure, r40_gehlhar2023safefl}. However, SMC introduces high communication overhead and remains vulnerable to inference attacks, as the final output can still leak information about individual inputs \cite{r41_truex2019hybrid}. Additionally, client dropouts in SMC lead to delays, reduced system robustness, and incomplete model updates. In FL, SMC protocols can be adapted to handle client dropouts by using redundancy and failover mechanisms, ensuring continued training by redistributing tasks to active clients. This improves model robustness, as adaptive client scheduling and reliability selection further mitigate the negative effects of dropouts \cite{r78_wang2022combating, r79_zawad2023hdfl}.

HE has emerged as an important technique in FL for preserving data privacy by enabling computations on encrypted data without revealing sensitive information. Despite its security advantages, HE often leads to increased computational and communication overhead, necessitating optimization strategies to maintain efficiency \cite{r64_madi2021secure, r65_park2022privacy, r66_shi2023privacy}. FPGA hardware accelerators technique, as demonstrated by Yang \textit{et al.} \cite{r67_yang2020fpga}, offer significant improvements in HE efficiency within FL systems. However, they did not provide an in-depth empirical analysis considering large-scale datasets and models. Other methods, such as the PFMLP framework \cite{r32_fang2021privacy} and the FastECDLP algorithm \cite{r61_tang2023solving} are significant advancements in homomorphic encryption and elliptic curve-based systems, respectively. However, both present certain limitations. The PFMLP framework faces performance overhead due to homomorphic encryption operations, including communication overhead during training, encryption and decryption key length, and key replacement frequency, which can affect training efficiency. Conversely, while the FastECDLP algorithm significantly improves decryption efficiency in EC-based AHE schemes, it provides limited insights into its effect on encryption performance and scalability for longer plaintexts. Furthermore, there is minimal discussion on potential security implications for encryption within this context. The FedML-HE \cite{r18_jin2023fedml} system offers selective parameter encryption to reduce overheads, though it still faces challenges when scaling to more extensive models. 

To further mitigate the overhead of HE in FL, quantization and sparsification techniques have been employed. BatchCrypt \cite{r38_zhang2020batchcrypt} and similar methods encode quantized gradients into compact representations, reducing encryption and communication costs while maintaining accuracy in cross-silo FL settings. However, these approaches can struggle to scale efficiently with larger deep learning models due to increased overhead. Zhu \textit{et al.}'s \cite{r62_zhu2021distributed} distributed additive encryption combined with quantization provides a notable reduction in computational demands by focusing on key parameters rather than encrypting the entire model. Techniques like DAEQ-FL enhance privacy in FL using additive ElGamal encryption and ternary quantization, but they face challenges in scaling to larger models due to the limited precision of ternary quantization, which can result in reduced accuracy for complex models \cite{r62_zhu2021distributed}. Sparsification techniques, such as those used in the FLASHE \cite{r37_jiang2021flashe} scheme and Ma \textit{et al.}'s \cite{r31_ma2022privacy} xMK-CKKS protocol, further enhance efficiency by reducing data volume and supporting modular operations, which involve performing computations within a fixed numerical range, thereby ensuring encryption consistency. This approach not only leads to lower communication costs but also improves computational performance. Despite these advancements, challenges remain in the deployment of HE in FL, primarily due to vulnerabilities to honest-but-curious clients and servers, as well as the inherent computational and communication bottlenecks during model training. While HE provides strong post-quantum security and preserves model performance by avoiding artificial noise and trusted environments, it still requires significant computational resources, especially in large-scale scenarios \cite{r61_tang2023solving, r23_xie2024efficiency}.

To summarize, each privacy-preserving technique in FL has its own strengths and limitations. SMC provides strong privacy by enabling secure data aggregation without revealing individual contributions, but it incurs high communication overhead and is vulnerable to inference attacks, especially with client dropouts. DP is straightforward to implement, but its addition of noise can degrade model accuracy, limiting its use in precision-critical tasks. HE offers robust privacy by allowing computations on encrypted data without compromising accuracy or requiring major algorithm changes. Although its computational and communication overheads have traditionally hindered its scalability for large-scale deployments, ongoing advancements in optimization techniques and hardware acceleration are addressing these challenges, making HE increasingly viable for real-time applications. One notable advancement is the use of quantization techniques, which reduce the precision of encrypted data to minimize communication costs while maintaining accuracy. Another key improvement is the introduction of hardware acceleration with GPUs and FPGAs, significantly boosting the performance of HE computations by offloading intensive operations, such as modular arithmetic, onto specialized hardware. In conclusion, the choice of privacy-preserving technique in FL must consider the specific use case, as trade-offs between privacy, performance, and scalability are critical. Techniques like quantization combined with model pruning or sparsification in the FHE process can significantly reduce communication, storage, and training costs, all while maintaining performance integrity in both cross-silo and cross-device applications.

%% file: Threa_model.tex
\section{Threat Model} \label{sec:ThreatModel}
In our mechanism, we assess the security of our method under two distinct adversarial scenarios. In the first scenario, the adversary has no knowledge of the model architecture, training parameters, or the underlying FL protocol. HE ensures that all communications between clients and the server remain fully encrypted, preventing the extraction of meaningful information even if encrypted model weights or parameters are intercepted, making this the strongest security assumption. In the second scenario, HE is applied, but the adversary gains access to decrypted model parameters or weights by compromising one or more participants, including honest but curious servers attempting to infer information. This simulates a more realistic threat model where adversaries leverage decrypted model parameters to infer sensitive data. To mitigate this, our framework employs pruning during FL, which significantly limits the adversary’s ability to reconstruct raw client data or infer sensitive information, even when partial or full access to decrypted parameters or weights.

\subsection{Overview of Threat Model}

FL has gained significant attention in both industry and academia due to its ability to train a global model across localized training data from multiple participants. However, this collaborative approach to machine learning is not without its vulnerabilities. MIA has been identified as a serious privacy concern in FL \cite{r5_zhang2020gan}. These attacks involve adversaries training classification models to determine if a data record is part of the model's training dataset, potentially leading to privacy breaches \cite{r5_zhang2020gan}. Unlike MIA, which aim to infer whether a specific data point was included in the training set, GIA \cite{r80_qian2024gi} go a step further by reconstructing the actual training data from the gradients exchanged during model updates. While MIA threatens privacy by revealing membership information, GIA poses a more direct risk by recovering sensitive input data, such as images or personal information. One specific type of inference attack is the user-level inference attack, which targets individual users participating in FL \cite{r68_zhao2021user}. Additionally, local model reconstruction attacks have been studied, where adversaries eavesdrop on messages exchanged between clients and servers to reconstruct personalized models \cite{r9_geiping2020inverting, r69_driouich2022local}.

\subsection{Mathematical Formulation} 

In this section, we focus on the security threats to FL, particularly on GIA \cite{r9_geiping2020inverting}. In a GIA, an adversary intercepts the gradients shared by clients with the central server and attempts to reconstruct the original input data by optimizing the difference in gradient directions. This optimization helps the adversary because the direction of the gradients provides information on how the model's parameters need to change to minimize the loss for specific input data. By iteratively adjusting a guessed input to align its gradient with that of the actual input, the adversary can gradually reconstruct the original data.

The adversary's goal can be formulated as an optimization problem. The equation below represents the objective for reconstructing client data by minimizing the cosine similarity between gradients. Cosine similarity is used because it measures the alignment of gradient directions, making it an ideal metric for determining how closely the reconstructed gradients match the actual ones, which helps refine the reconstruction process.

\begin{equation}
\arg \min_{x \in [0,1]^n} \left( 1 - \frac{\langle \nabla_\theta L_\theta(x, y), \nabla_\theta L_\theta(x^*, y) \rangle}{\|\nabla_\theta L_\theta(x, y)\| \, \|\nabla_\theta L_\theta(x^*, y)\|} \right) + \alpha \cdot \text{TV}(x) \label{eq1recon}
\end{equation}

In Equation \ref{eq1recon}, the gradient of the loss function with respect to the model parameters for input \( x \), denoted as \( \nabla_\theta L_\theta(x, y) \), is compared to the gradient for the ground truth input \( x^* \), \( \nabla_\theta L_\theta(x^*, y) \). The objective focuses on aligning the gradient directions rather than their magnitudes, as the direction of the gradient provides crucial information about how the model's predictions change in response to specific input data. By aligning the gradient directions, the adversary ensures that the reconstructed input leads to similar changes in the model’s loss function as the original input, making the reconstruction more accurate. The gradient magnitudes, on the other hand, primarily reflect how far the model is from optimality, which can vary during training and does not contribute significantly to reconstructing the input. To preserve image smoothness, a regularization term \( \alpha \cdot \text{TV}(x) \), where \( \text{TV}(x) \) represents the total variation of the image, is included. Total Variation (TV) is commonly used in image reconstruction to reduce noise while preserving important edges in the image. By incorporating \( \text{TV}(x) \) in the optimization, the adversary ensures that the reconstructed image remains visually plausible, with reduced noise and preserved critical structures.

\subsection{Attack Practicality}
Inference attacks, particularly gradient inversion attacks, have been demonstrated to pose significant and practical threats to FL systems. In these attacks, adversaries exploit the gradients or model updates exchanged between clients and the central server to reconstruct sensitive training data. Gradients are used to optimize model parameters by representing how much and in what direction the model’s predictions should change with respect to the input data. However, this same process can unintentionally expose information about the underlying data because the gradients encode key features of the input, such as patterns or specific attributes that the model learns from. By analyzing these gradients, attackers can reverse-engineer and reconstruct the original input data, especially in models where the gradient directions reveal detailed information about the features that influence the model's predictions. This makes gradient inversion attacks both feasible and dangerous in real-world settings, as the gradients provide a rich source of data that adversaries can exploit.  Connected autonomous vehicles (CAVs)  can benefit from FL to allow for decentralized decision-making.  CAV applications can be vulnerable to adversarial attacks during both the FL training and inference phases \cite{mia2024secure}. FL allows AVs to improve models collaboratively by sharing aggregated model without exposing personal information. However, attacks like GIA and MIA still pose serious risks, as shared data may contain sensitive information about drivers, vehicle routes, and surroundings. For example, GIA could reconstruct images or AV sensor data collected during training, potentially revealing private information about pedestrians and drivers. These privacy threats are significant, as exposed data could disclose personal information or allow attackers to track specific vehicles by analyzing patterns.

Studies, such as \textit{MiBench}, provide substantial evidence of this vulnerability by showing that even the partial extraction of gradient information can lead to the revelation of highly sensitive personal data, such as medical or financial records, in real-time environments \cite{r75_qiu2024mibench}. Gradients used for model optimization encode important features of the input data that are critical to the model’s predictions. As a result, even small portions of gradient data can reveal patterns or attributes from the original dataset, which can be exploited by attackers to reconstruct sensitive information. This illustrates that these attacks are not merely theoretical but are actively being used to extract private information from models during deployment.

In another real-world example, adversaries used generative AI models to extract sensitive data from shared model outputs. In FL, the attack occurs when an adversary exploits the gradients shared by a victim during the training process to train a GAN (Generative Adversarial Network). The GAN’s generator produces synthetic data that mimics the victim's private dataset, while the discriminator evaluates the quality based on the gradients. Over several iterations, the GAN learns to generate data closely resembling the victim's original data, allowing the attacker to reconstruct sensitive information and reverse-engineer the data distribution \cite{r81_zhang2020exploiting}. This further underscores the practicality and danger of inference attacks, especially in large-scale FL deployments \cite{r76_chen2024private}. These instances highlight the need for enhanced defenses like HE to protect FL systems from evolving threats. HE encrypts gradients during training, preventing adversaries and even the honest but curious server from accessing raw gradient data, thus preventing gradient inversion attacks. Additionally, gradient pruning limits the amount of gradient information exchanged, further reducing vulnerabilities. Together, these techniques significantly mitigate the risks of inference attacks during FL training and inference phase.

%% file: Methodology.tex
\section{Methodology} \label{sec:Methodology}

\begin{figure*} [!t]
    \centering
    \includegraphics[width=0.9 \linewidth]{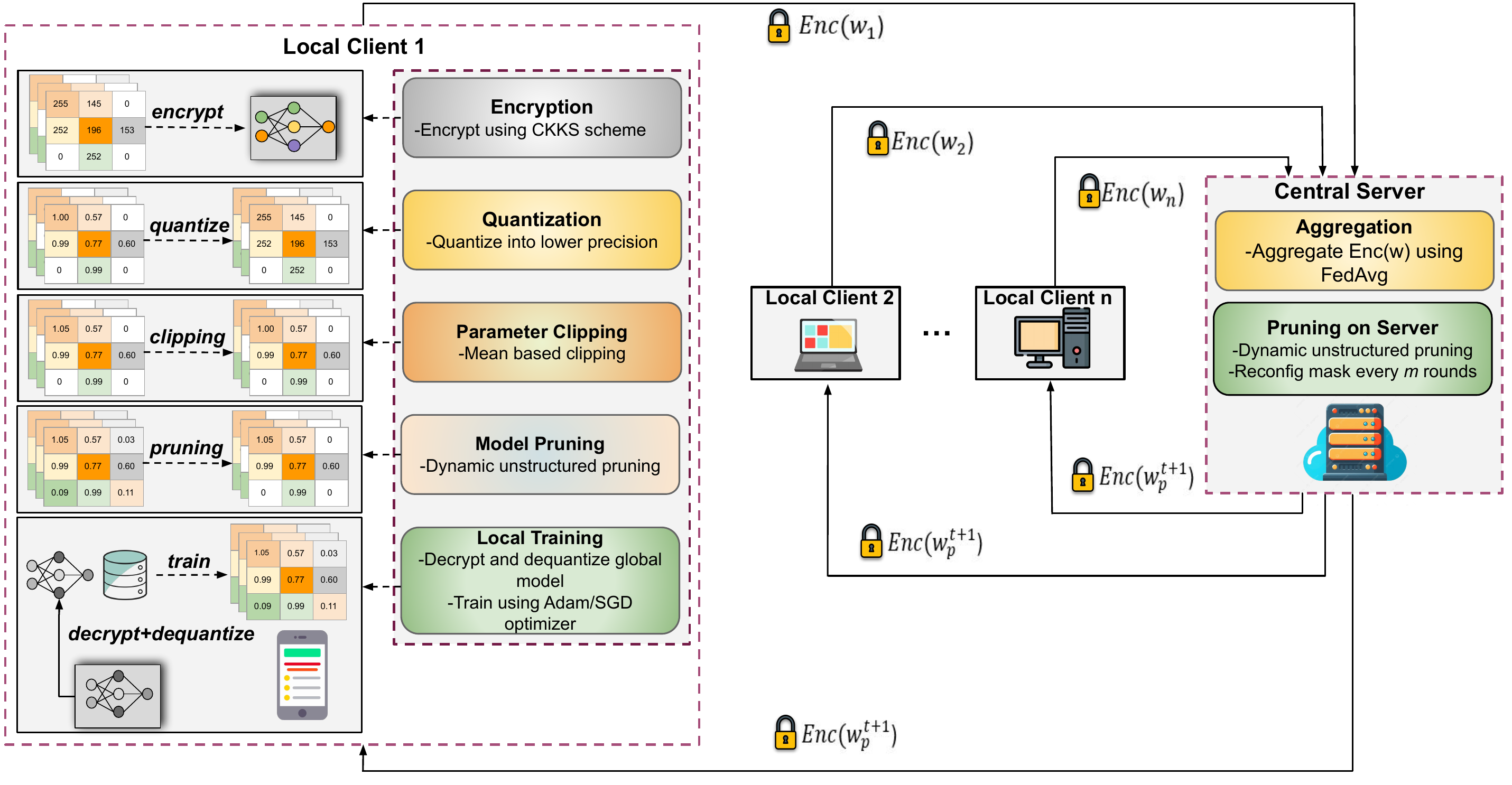}
    \caption{Overview of the proposed QuanCrypt-FL framework.}
    \label{fig:figure_1}
\end{figure*}

In our proposed method, we implemented Homomorphic Encryption (HE) in FL with quantization and pruning to enhance training efficiency, reduce communication costs, and make the training process more resilient against inference attacks. QuanCrypt-FL framework is visualized in Figure \ref{fig:figure_1}.

% first figure
The process begins by sharing the initial global model parameters \( \mathbf{w}^0 \) with all clients. Each client performs local training on its own dataset, followed by the aggregation of model updates at the server.

For each communication round \( t \), the server first sends the current global model parameters \( \mathbf{w}^t \) to each client \( C_i \). The clients then perform local training using their respective datasets \( D_i \). During local training, each client computes the model update by minimizing the loss on its local dataset. The model update is calculated using the equation \ref{eq:localtrain}.

\begin{equation}
\Delta \mathbf{w}_i^{t+1} = \mathbf{w}^t - \eta \nabla L(\mathbf{w}^t, D_i)
\label{eq:localtrain}
\end{equation}

\noindent where \( \mathbf{w}^t \) are the global model parameters shared by the server at round \( t \), \( \Delta \mathbf{w}_i^{t+1} \) represents the model update for client \( i \) after local training, \( \eta \) is the learning rate used by the optimizer, and \( L(\mathbf{w}^t, D_i) \) is the loss function evaluated on the local dataset \( D_i \).

During local model training, we employed a pruning technique to iteratively remove less important weights or gradients from the model updates. Specifically, clients perform soft, unstructured pruning based on the L1 norm, which creates a sparse model and makes the FL training process more efficient. The pruning process is guided by a dynamically updated pruning rate \( p_t \), which increases over the communication rounds, allowing for more aggressive pruning as training progresses. After pruning, clients send their pruned updates to the server, which aggregates them using FedAvg to generate the global model. This pruning technique not only reduces the model size and computational costs but also makes the training process more resistant to inference attacks. 

\begin{figure}[htp!]
    \centering
    \includegraphics[width=0.95\linewidth]{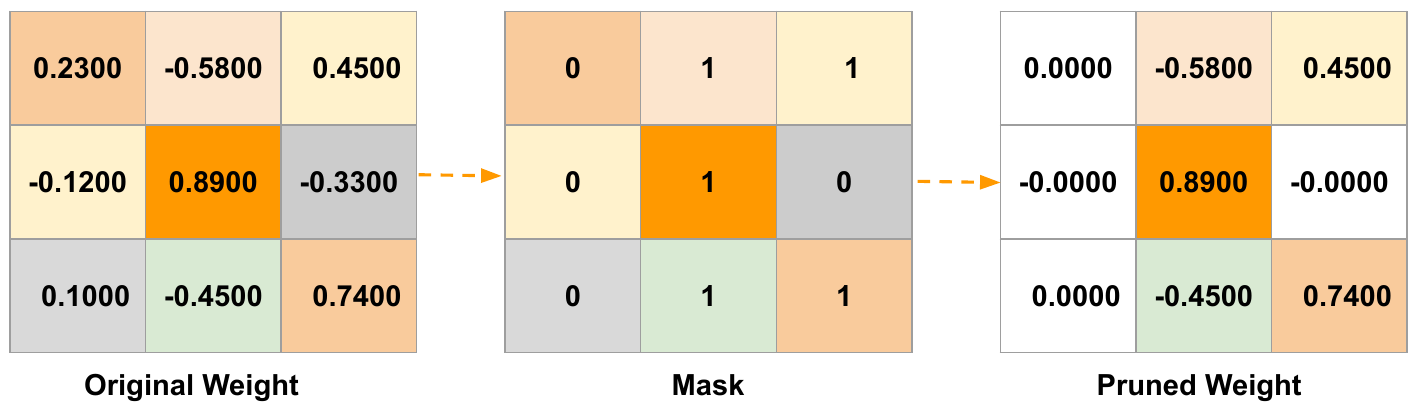}
    \caption{Model sparsification using unstructured pruning based on L1 norm.}
    \label{figure_prune}
\end{figure}

By progressively increasing the pruning rate, the communication efficiency improves throughout the rounds. As clients share a sparsified model with the server, the transmitted model is no longer the full model, limiting the information available to potential attackers. The sparsity introduced by pruning constrains the parameter space, significantly reducing the chances of reverse engineering or inferring sensitive data. This reduction in exposed parameters inherently enhances privacy protection, making it more difficult for adversaries to extract meaningful insights about the underlying data. Weight pruning or sparsification process is visualized in Figure \ref{figure_prune}.

The pruning rate \( p_t \) is updated iteratively using the  equation \ref{eq:prate}.

\begin{equation}
p_t = \max\left(0, \frac{t - t_{\text{eff}}}{t_{\text{target}} - t_{\text{eff}}}\right) \times \left(p_{\text{target}} - p_0\right) + p_0
\label{eq:prate}
\end{equation}

\noindent where \( p_t \) is the pruning rate at round \( t \), \( t_{\text{eff}} \) is the effective round when pruning starts, \( t_{\text{target}} \) is the target round when the target pruning rate is reached, \( p_0 \) is the initial pruning rate, and \( p_{\text{target}} \) is the target pruning rate. This pruning rate increases gradually from the initial value to the target value, ensuring that pruning is progressively applied more aggressively as training advances.

Once pruning is applied to the model updates at each client, the pruned local model update \( \Delta \mathbf{w}_{p,i}^{t+1} \) is computed as in equation \ref{eq:localupdate}.

\begin{equation}
\Delta \mathbf{w}_{p,i}^{t+1} = \Delta \mathbf{w}_i^{t+1} \odot m_i^{t}
\label{eq:localupdate}
\end{equation}

\noindent where \( \odot \) represents the element-wise product, and \( m_i^{t} \) is the local pruning mask generated to identify which weights to prune at communication round \( t \). This pruned update \( \Delta \mathbf{w}_{p,i}^{t+1} \) is then quantized and sent to the server for aggregation.
\RestyleAlgo{algoruled}
\setcounter{AlgoLine}{0}
\begin{algorithm}[ht]
\SetAlgoLined
\KwIn{model update \( \Delta w_i^{t+1} \), clip factor \( \alpha \)}
\KwOut{Clipped model update \( \Delta w_{\text{C}, i}^{t+1} \)}

\ForEach{parameter \( \Delta w_i^{t+1} \) in the model update}{
    \If{\( \Delta w_i^{t+1} \) is not floating-point}{
        Convert \( \Delta w_i^{t+1} \) to floating-point
    }
    Compute \( \mu_i = \text{mean}(|\Delta w_i^{t+1}|) \)
    
    \( \Delta w_{\text{C}, i}^{t+1} \gets \mathit{clip}(\Delta w_i^{t+1}, -\alpha \cdot \mu_i, \alpha \cdot \mu_i) \)
}

\Return \( \Delta w_{\text{C}, i}^{t+1} \)

\caption{Dynamic Mean-Based Clipping Technique}
\label{algo1}
\end{algorithm}

\begin{figure*} [!t]
    \centering
    \includegraphics[width=1.0 \linewidth]{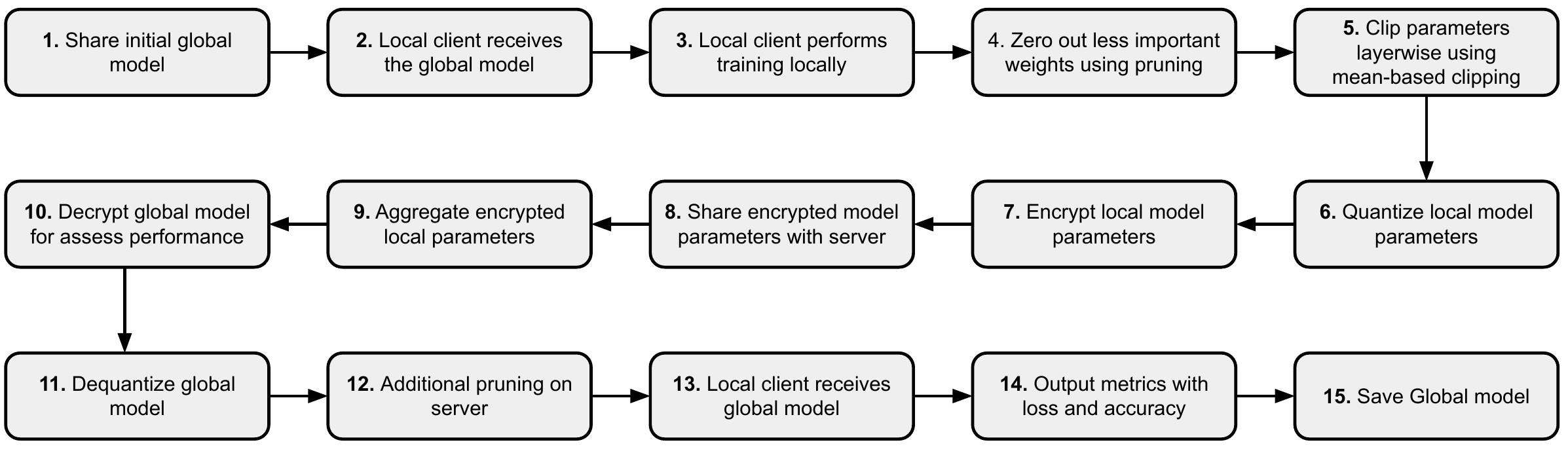}
    \caption{High Level Overview of the proposed QuanCrypt-FL mechanism.}
    \label{quantphefl_highlevel}
\end{figure*}

We also employed a dynamic mean-based layer-wise clipping technique in Algorithm \ref{algo1} to help reduce inconsistencies during the training process. The clipping factor controls the clipping parameter, dynamically adjusting the clipping based on layer-wise updates, rather than using a static clipping method. This approach ensures that each layer's updates are clipped according to their specific dynamics, leading to more stable and efficient training. After the local model updates are computed, each client clips its own model update \( \Delta \mathbf{w}_i^{t+1} \) to avoid instability before sending it to the server. The clipping for client \( i \)'s model update is applied using the following equation \ref{eq:clipping}.

\begin{equation}
\Delta \mathbf{w}_{\text{C}, i}^{t+1} = \mathit{clip} \left( \Delta \mathbf{w}_i^{t+1}, -\alpha \cdot \mu_i, \alpha \cdot \mu_i \right)
\label{eq:clipping}
\end{equation}

\noindent where \( \mu_i \) is the mean of the absolute values of the elements of client \( i \)'s model update, calculated as in equation \ref{eq:mean} .

\begin{equation}
\mu_i = \frac{1}{n} \sum_{j=1}^{n} \left|\Delta \mathbf{w}_{i, j}^{t+1}\right|
\label{eq:mean}
\end{equation}

\noindent The clipping function \( \mathit{clip}(\Delta \mathbf{w}_i^{t+1}, a, b) \) ensures that the values of \( \Delta \mathbf{w}_i^{t+1} \) are constrained within the range \( [-\alpha \cdot \mu_i, \alpha \cdot \mu_i] \), limiting the impact of extreme values.

Next, each client performs quantization on the pruned and clipped model updates to reduce communication costs. The quantization process involves calculating the scaling factor \( s \) and determining the quantized values \( q_x \) to ensure the updates are compressed before transmission. The scaling factor \( s \) is calculated using equation \ref{eq:quantscale}.

\begin{equation}
s = 
\begin{cases} 
\frac{1.0}{q_{\text{max}} - q_{\text{min}}}, & \text{if } x_{\text{max}} = x_{\text{min}} = 0 \\
\frac{x_{\text{min}}}{q_{\text{max}} - q_{\text{min}}}, & \text{if } x_{\text{max}} = x_{\text{min}} \neq 0 \\
\frac{x_{\text{max}} - x_{\text{min}}}{q_{\text{max}} - q_{\text{min}}}, & \text{otherwise}
\end{cases}
\label{eq:quantscale}
\end{equation}

\noindent Using this scaling factor \( s \), the quantized values \( q_x \) are then computed as in equation \ref{eq:qscaling}.

\begin{equation}
q_x = \mathit{round} \left( \frac{x - x_{\text{min}}}{s} + z_0 \right)
\label{eq:qscaling}
\end{equation}

\noindent where \( z_0 \) represents the zero-point, calculated as:\\ $
z_0 = \mathit{clip} \left( q_{\text{min}}, q_{\text{max}}, q_{\text{min}} - \frac{x_{\text{min}}}{s} \right)
$.

\noindent The quantized values \( q_x \) are then clamped to the range \([q_{\text{min}}, q_{\text{max}}]\) and converted to the appropriate data type based on the bit width (e.g., 8-bit, 16-bit, or 32-bit) to minimize communication overhead.

After completing quantization, each client encrypts the quantized model updates using the CKKS homomorphic encryption scheme. The server then aggregates the encrypted updates from all clients using the following equation: $
\mathit{AggEnc}(\Delta \mathbf{w}_q) = \frac{1}{N} \sum_{i=1}^{N} \mathit{Enc}(\Delta \mathbf{w}_{q, i}^{t+1})$. \\
Leveraging the homomorphic capabilities of CKKS, the server performs this aggregation directly on the encrypted model updates, without requiring decryption of individual updates. Since the server operates solely on ciphertexts, it can compute the sum of the encrypted updates element-wise while keeping each client’s data private. This process ensures that even during aggregation, the server has no access to the underlying data. The resulting aggregated model remains encrypted and can be decrypted only by a trusted party or the clients, thus preserving data privacy throughout the FL training process. This approach aligns with the security and efficiency objectives of FL by enabling secure computations while safeguarding client data.\\
After aggregation, the server decrypts the global model update to evaluate its performance using the following equation: $\Delta \mathbf{w}_{q} = \mathit{Dec}\left( \mathit{Enc}(\Delta \mathbf{w}_q) \right)$.  While the server can assess the overall model, individual client updates remain encrypted, ensuring privacy is preserved. This approach balances the need for performance evaluation with the protection of client data in the FL process.\\
The server then performs dequantization to recover the floating-point values using the following equation:

$ x' = s \cdot (q_x - z_0). $

After the global model is updated and dequantized, pruning is applied to eliminate certain weights based on pruning rate. The pruned model weights \( w_p^{t+1} \) are obtained by applying the pruning mask \( m_{t} \) to the global model weights \( w_{\text{dq}}^{t+1} \) as stated in here: 

$w_p^{t+1} = w_{\text{dq}}^{t+1} \odot m_{t} $,

 where \( \odot \) represents the element-wise (Hadamard) product, and \( m_{t} \) is the pruning mask applied to eliminate certain weights in the global model. 
Once pruning is applied to the global model, the pruned global model \( w_p^{t+1} \) is sent to the clients for the next communication round. The clients will apply the same process again: local training, pruning, clipping, quantization, and secure aggregation. The pruning rate \( p_t \) is updated iteratively for each round, and the process continues until the model converges. The final QuanCrypt-FL mechanism is presented in Algorithm \ref{algo2}. Moreover, step by step procedure of our method is depicted in Figure \ref{quantphefl_highlevel}.

\RestyleAlgo{algoruled}
\setcounter{AlgoLine}{0}
\begin{algorithm}[ht]
\SetAlgoLined
\KwIn{Global model \( w^0 \), clients \( N \), rounds \( T \), learning rate \( \eta \), clip factor \( \alpha \), HE context \( \mathcal{H} \), quantization bit-length \( q_{\text{bit}} \), initial pruning rate \( p_0 \), target pruning rate \( p_{\text{target}} \), scaling factor \( s \), zero-point \( z_0 \), quantized weights \( \Delta w_q \), dequantized weights \( w_{\text{dq}} \), pruned weights \( w_p \)}
\KwOut{Trained global model \( w^T \)}

\textbf{Init:} \( w^0 \), \( \mathcal{H} \), client datasets, effective round \( t_{\text{eff}} \), target round \( t_{\text{target}} \)

\For{each round \( t = 1, \dots, T \)}{
    Server sends \( w^t \) to all clients
    
    \For{each client \( i = 1, \dots, N \)}{
        \(\Delta w_i^{t+1} \gets w^t - \eta \nabla L(w^t, D_i) \)
        
        \( p_t \gets \max\left(0, \frac{t - t_{\text{eff}}}{t_{\text{target}} - t_{\text{eff}}}\right) \times \left(p_{\text{target}} - p_0\right) + p_0 \)
        
        \( m_i^{t} \gets \mathit{pruning\_mask}(\Delta w_i^{t+1}, p_t, L1\_norm) \)
        
        \( \Delta w_{p,i}^{t+1} \gets \Delta w_i^{t+1} \odot m_i^{t} \)
        
        \( \Delta w_{\text{C}, i}^{t+1} \gets \mathit{clip}(\Delta w_{p,i}^{t+1}, -\alpha \cdot \mu_i, \alpha \cdot \mu_i) \)
        
        \( s \gets 
        \begin{cases} 
        \frac{1.0}{q_{\text{max}} - q_{\text{min}}}, & x_{\text{max}} = x_{\text{min}} = 0 \\
        \frac{x_{\text{min}}}{q_{\text{max}} - q_{\text{min}}}, & x_{\text{max}} = x_{\text{min}} \neq 0 \\
        \frac{x_{\text{max}} - x_{\text{min}}}{q_{\text{max}} - q_{\text{min}}}, & \text{otherwise}
        \end{cases}
        \)
        
        \( q_x \gets \mathit{round} \left( \frac{\Delta w_{\text{C}, i}^{t+1} - x_{\text{min}}}{s} + z_0 \right) \)
        
        \( q_x \gets \mathit{clip}(q_x, q_{\text{min}}, q_{\text{max}}) \)
        
        \( \Delta w_{q, i}^{t+1} \gets \mathit{Enc}(q_x, \mathcal{H}) \)
    }
    
    \( w_{\text{agg}}^{t+1} \gets \frac{1}{N} \sum_{i=1}^{N} \Delta w_{q, i}^{t+1} \)
    
    \( w^{t+1} \gets \mathit{Dec}(w_{\text{agg}}^{t+1}, \mathcal{H}) \)
    
    \( w_{\text{dq}}^{t+1} \gets s \cdot (w^{t+1} - z_0) \)
    
    \( m_t \gets \mathit{pruning\_mask}(w_{\text{dq}}^{t+1}, p_t, L1\_norm) \)
    
    \( w_p^{t+1} \gets w_{\text{dq}}^{t+1} \odot m_{t} \)
    
    Server sends \( w_p^{t+1} \) to clients
}

\caption{QuanCrypt-FL}
\label{algo2}
\end{algorithm}

%% file: Experiments_and_result_analysis.tex
\section{Experiments and Result Analysis} \label{sec:experiments}

\subsection{Experimental Setup}

Our experiments were conducted on an Ubuntu server equipped with two NVIDIA RTX A6000 GPUs, an Intel Core i9 processor, and 128 GB of RAM. We explored FL across 10 to 50 clients using both IID and non-IID data distribution strategies. Each client independently trained their local models on a subset of data, using the Adam optimizer with a learning rate of 0.001 and a weight decay of 0.0001. Batch sizes were set at 64, or 128 depending on the number of clients, and each client trained for one local epoch per communication round. To ensure data privacy and security, we implemented HE with TenSEAL \cite{r74_benaissa2021tenseal}, based on Microsoft SEAL, adopting the CKKS scheme with a polynomial modulus degree of 16384 and coefficient modulus sizes [60, 40, 40, 40, 60]. The encryption context contained a public and private key pair shared among all clients, with the server using only the public key to securely aggregate model updates without revealing individual client data. For evaluation, model weights were decrypted with the private key at the server when necessary to assess testing accuracy.

\subsection{Datasets and Models}

In our FL experiments, we used three datasets: CIFAR10 \cite{r70_krizhevsky2009learning}, CIFAR100 \cite{r70_krizhevsky2009learning}, and MNIST \cite{r71_lecun1998mnist}. These datasets were partitioned among clients using both IID and non-IID strategies to simulate different real-world data distribution scenarios. In the IID setting, data was evenly and randomly distributed among clients, whereas in the non-IID setting, each client received data containing only a subset of classes. CIFAR10 and CIFAR100 images were normalized using their respective mean and standard deviation values, while MNIST’s grayscale images were normalized accordingly. The training datasets were split into 80\% for training and 20\% for validation, with each client receiving a portion for local training. For evaluation purposes, the full test dataset from each respective dataset (CIFAR10, CIFAR100, and MNIST) was used, ensuring that the model performance was assessed on a standardized and unchanged test set after each communication round.

Several models were utilized in the experiments, including CNN, AlexNet, and ResNet18, with each model trained locally on client data to evaluate performance under IID data distribution. The is a CNN designed for MNIST dataset, consisting of two convolutional layers with ReLU activation, max-pooling, a fully connected layer, dropout, and an output layer. We modified AlexNet, adapted from the original AlexNet, includes four convolutional layers with 3x3 kernels, ReLU activations, max-pooling, dropout, two fully connected layers, and a softmax activation for classification. The ResNet18 model follows the standard ResNet-18 architecture, using BasicBlock to create four stages with increasing feature map sizes and downsampling, ending with average pooling and a fully connected output layer.

\subsection{Quantization and Pruning}
% Table

In our FL experiments, we employed quantization to compress model updates, with the option to use 8-bit, 16-bit, or 32-bit quantization, controlled by the quantization bit-length \( q_{\text{bit}} \), to reduce communication overhead. A clipping factor \( \alpha = 3.0 \) was applied to manage extreme values in the model updates, ensuring stable training. For pruning, the initial pruning rate \( p_0 \) was set to 20\%, and pruning began at the effective round \( t_{\text{eff}} = 40 \). The pruning rate increased progressively until it reached the target pruning rate \( p_{\text{target}} = 50\% \) by round \( t_{\text{target}} = 300 \). A pruning mask \( m_t \) was applied to generate pruned weights \( w_p \), reducing the model size and computational cost while maintaining accuracy.

\subsection{Clipping, Smoothing, and Checkpoints}

To control the magnitude of model updates, we applied a clipping mechanism with a clip factor \(\alpha\), set by default to 3.0 after conducting a grid search within the range \([1.0, 5.0]\). This ensured that extreme values in model updates were controlled, ensuring stability during training and preventing large deviations.

\begin{figure*} [htp!]
  \begin{minipage}[b]{0.5\linewidth}
    \centering
    \includegraphics[width=\linewidth]{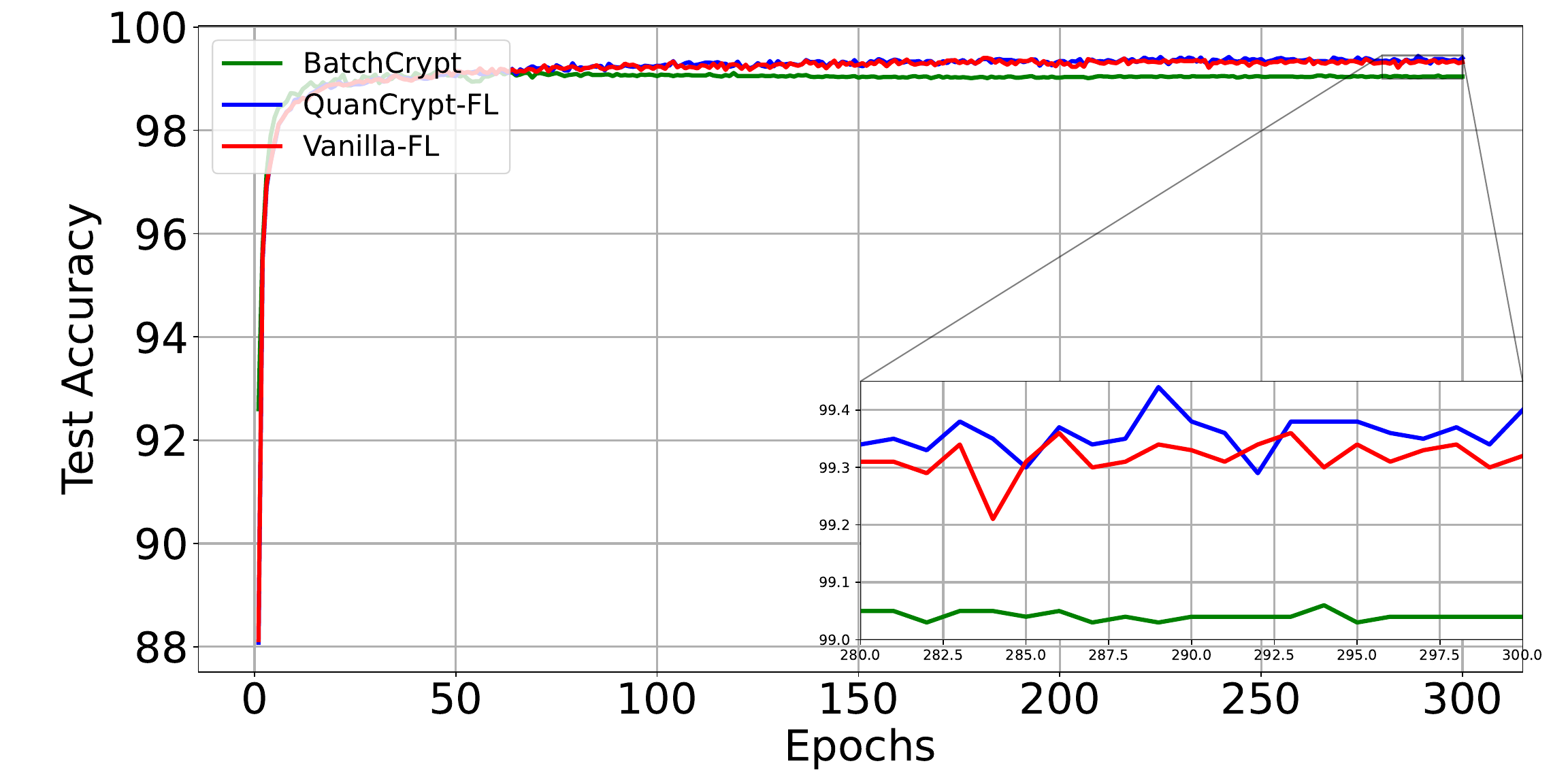}
    \caption{Comparison of methods considering 10 clients, Model: CNN, Dataset: MNIST, $\alpha$=3.0, $\lambda$=1.0.}
    \label{fig:mnist_c10}
  \end{minipage}
  \hspace{10pt} 
  \begin{minipage}[b]{0.5\linewidth}
    \centering
    \includegraphics[width=\linewidth]{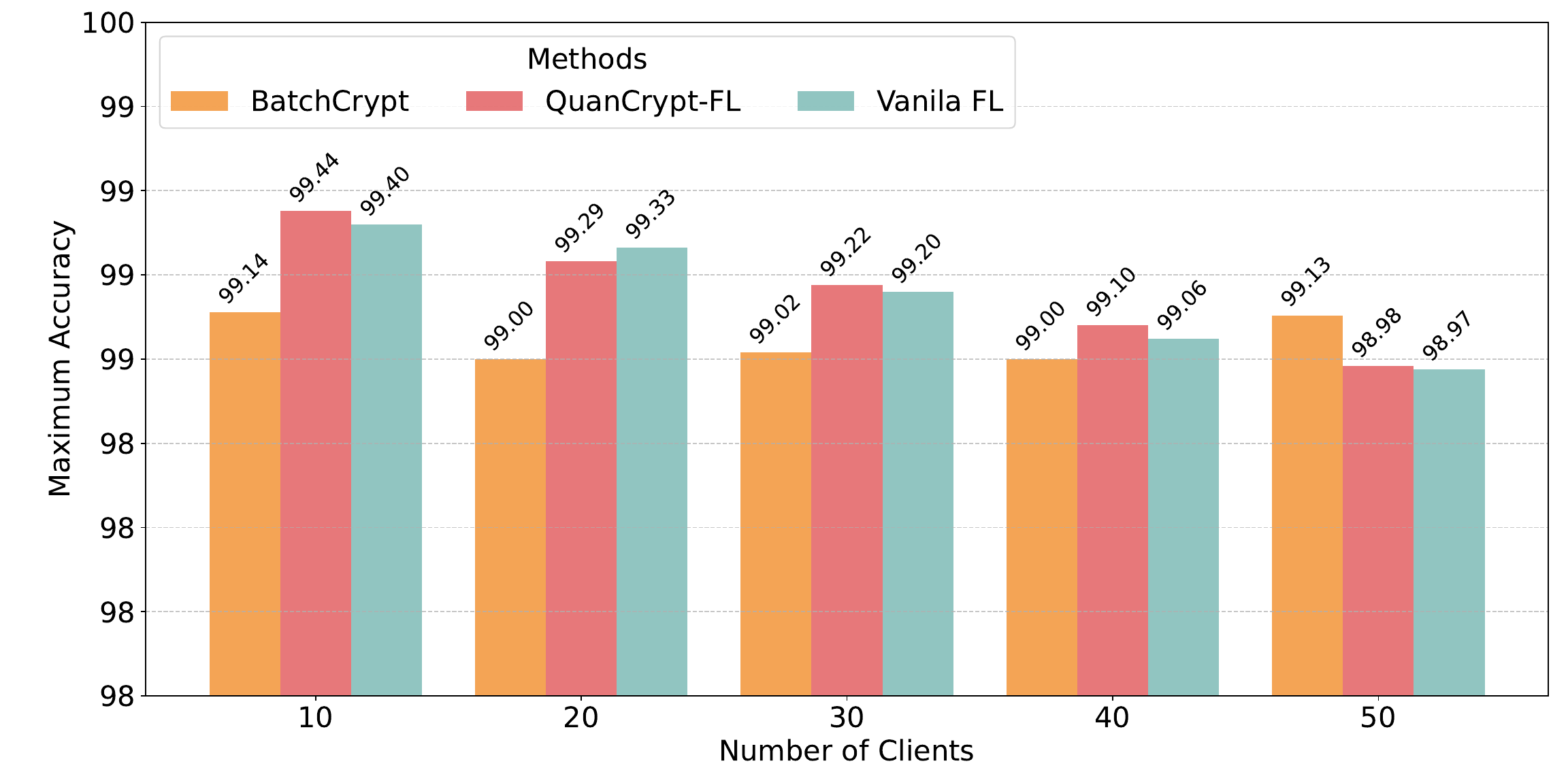}
    \caption{Comparison of Methods for several clients, Model: CNN, Dataset: MNIST, $\alpha$=3.0, $\lambda$=1.0.}
    \label{fig:mnist_cbar}
  \end{minipage}

\end{figure*}

In our FL model, we also incorporated a global learning rate hyperparameter \(\lambda\) to adjust the influence of new model updates. At each communication round, the initial global parameter vector \(\mathbf{w}_{t,\text{init}}\) and the updated parameter vector \(\mathbf{w}_{t,\text{final}}\) were combined according to the following update rule \cite{r73_salehi2024corbin}:

\[
\mathbf{w}_{t} = (1 - \lambda)\mathbf{w}_{t,\text{init}} + \lambda\mathbf{w}_{t,\text{final}}
\]

\noindent When \(\lambda = 1\), the equation reverts to the standard update without smoothing. A grid search over \(\lambda \in \{0.1, 0.2, \dots, 1\}\) was performed to select the best value for optimizing the model's convergence.

To further enhance model performance, we implemented a checkpointing mechanism with a patience of five epochs. If the validation accuracy did not improve after five consecutive epochs, the model was reloaded from the last checkpoint, ensuring that the model did not overfit or degrade in performance during training.

\subsection{Result Analysis}

We will provide a comparative analysis of our proposed method with BatchCrypt and Vanilla FL. The results will be evaluated across multiple metrics, including model accuracy, training time, encryption time, decryption time, inference time, and storage efficiency. This analysis will demonstrate the effectiveness of our approach in enhancing both computational performance and security in FL.

Figure \ref{fig:mnist_c10} shows the test accuracy comparison of BatchCrypt, QuanCrypt-FL, and Vanilla-FL on the MNIST dataset using a CNN model with 10 clients over 300 communication rounds. The results indicate that BatchCrypt starts with reasonably high accuracy but quickly stabilizes at around 99.04\%, consistently underperforming compared to both QuanCrypt-FL and Vanilla-FL. QuanCrypt-FL, however, closely tracks Vanilla-FL throughout the training, reaching an accuracy of 99.40\% by the final round, while Vanilla-FL achieves a similar result of 99.32\%. This minimal difference between QuanCrypt-FL and Vanilla-FL demonstrates that QuanCrypt-FL achieves nearly identical performance to Vanilla-FL, while offering additional privacy-preserving features. BatchCrypt, although competitive, lags behind both methods across all 300 rounds, showing the trade-offs involved in using a heavier encryption mechanism. Ultimately, QuanCrypt-FL maintains strong accuracy comparable to Vanilla-FL and clearly outperforms BatchCrypt, making it a more effective choice when both privacy and accuracy are essential.

Figure \ref{fig:mnist_cbar} illustrates the maximum accuracy achieved by BatchCrypt, QuanCrypt-FL, and Vanilla-FL across different client counts (10, 20, 30, 40, 50) on the MNIST dataset using a CNN model. QuanCrypt-FL consistently performs at the highest level, achieving 99.44\% for 10 clients, slightly outperforming Vanilla-FL at 99.40\%. As the number of clients increases, QuanCrypt-FL maintains strong accuracy with 99.29\% for 20 clients, 99.22\% for 30 clients, 99.10\% for 40 clients, and 98.98\% for 50 clients. Vanilla-FL closely follows with accuracies of 99.33\%, 99.20\%, 99.06\%, and 98.97\% for the same client counts, respectively. In contrast, BatchCrypt consistently underperforms compared to both QuanCrypt-FL and Vanilla-FL, reaching 99.14\% for 10 clients and showing slightly lower values for the remaining client counts, with accuracies ranging between 99.00\% and 99.13\%. While BatchCrypt provides competitive results, it consistently lags behind the higher performance of QuanCrypt-FL and Vanilla-FL, demonstrating the superior ability of QuanCrypt-FL to maintain high accuracy while incorporating privacy-preserving features.

\begin{figure*} [htp!]
  \begin{minipage}[b]{0.5\linewidth}
    \centering
    \includegraphics[width=\linewidth]{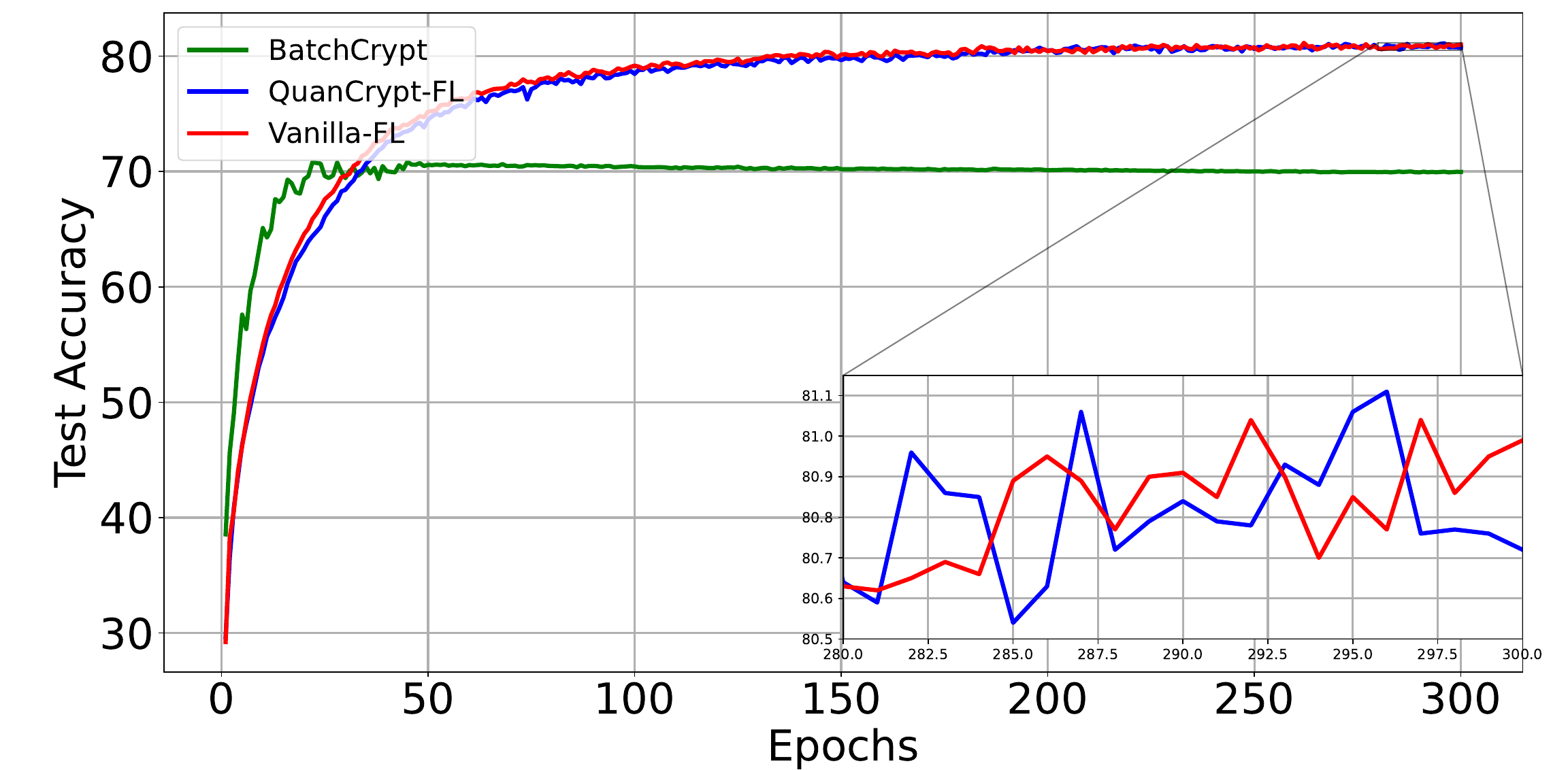}
    \caption{Comparison of accuracy considering 10 clients, Model: AlexNet, Dataset: CIFAR10, $\alpha$=3.0, $\lambda$=1.0.}
    \label{fig:cifar10alxc10}
  \end{minipage}
  \hspace{10pt} 
  \begin{minipage}[b]{0.5\linewidth}
    \centering
    \includegraphics[width=\linewidth]{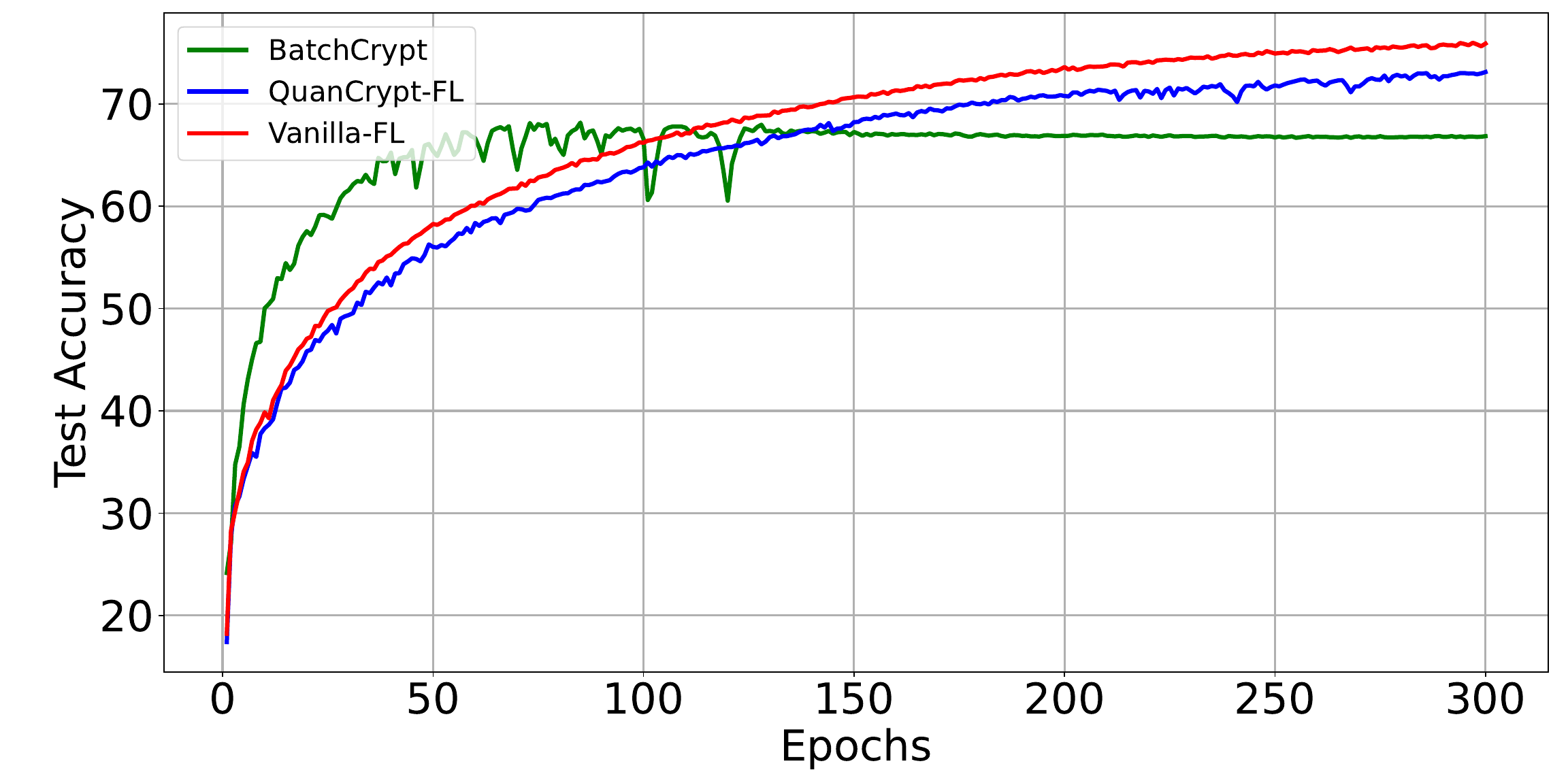}
    \caption{Comparison of accuracy considering 50 clients, Model: AlexNet, Dataset: CIFAR10, $\alpha$=3.0, $\lambda$=1.0.}
    \label{fig:cifar10alxc50}
  \end{minipage}

  \label{fig:comparison_clients}
\end{figure*}

A comparison of BatchCrypt, QuanCrypt-FL, and Vanilla-FL on the CIFAR-10 dataset with 10 clients reveals key differences in performance, as shown in Figure \ref{fig:cifar10alxc10}. While BatchCrypt initially achieves slightly higher accuracy in the early rounds, it quickly falls behind as training progresses. By the final epochs, QuanCrypt-FL and Vanilla-FL reach similar accuracy levels of approximately 80.40\%, whereas BatchCrypt lags significantly with a final accuracy of 69.96\%. This substantial performance differ of over 10.00\% underscores the trade-offs inherent in BatchCrypt, where the pursuit of privacy leads to a notable compromise in accuracy. On the other hand, QuanCrypt-FL closely matches the performance of Vanilla-FL, demonstrating its ability to maintain accuracy while offering including privacy features.

In the case of 50 clients, the differences in performance between BatchCrypt, QuanCrypt-FL, and Vanilla-FL become even more pronounced, as shown in Figure \ref{fig:cifar10alxc50}. While BatchCrypt initially shows some promise in the early rounds, it fails to keep up as training progresses. By the final epochs, QuanCrypt-FL achieves an accuracy of 73.12\%, closely aligning with Vanilla-FL's 75.90\%, while BatchCrypt trails behind with 66.85\%. This (7-9)\% difference further highlights BatchCrypt’s limitations in scaling effectively. In contrast, QuanCrypt-FL consistently maintains competitive accuracy, making it a reliable and scalable choice for FL scenarios with larger client bases.

\begin{figure*} [!t]
    \centering
    \includegraphics[width=0.9 \linewidth]{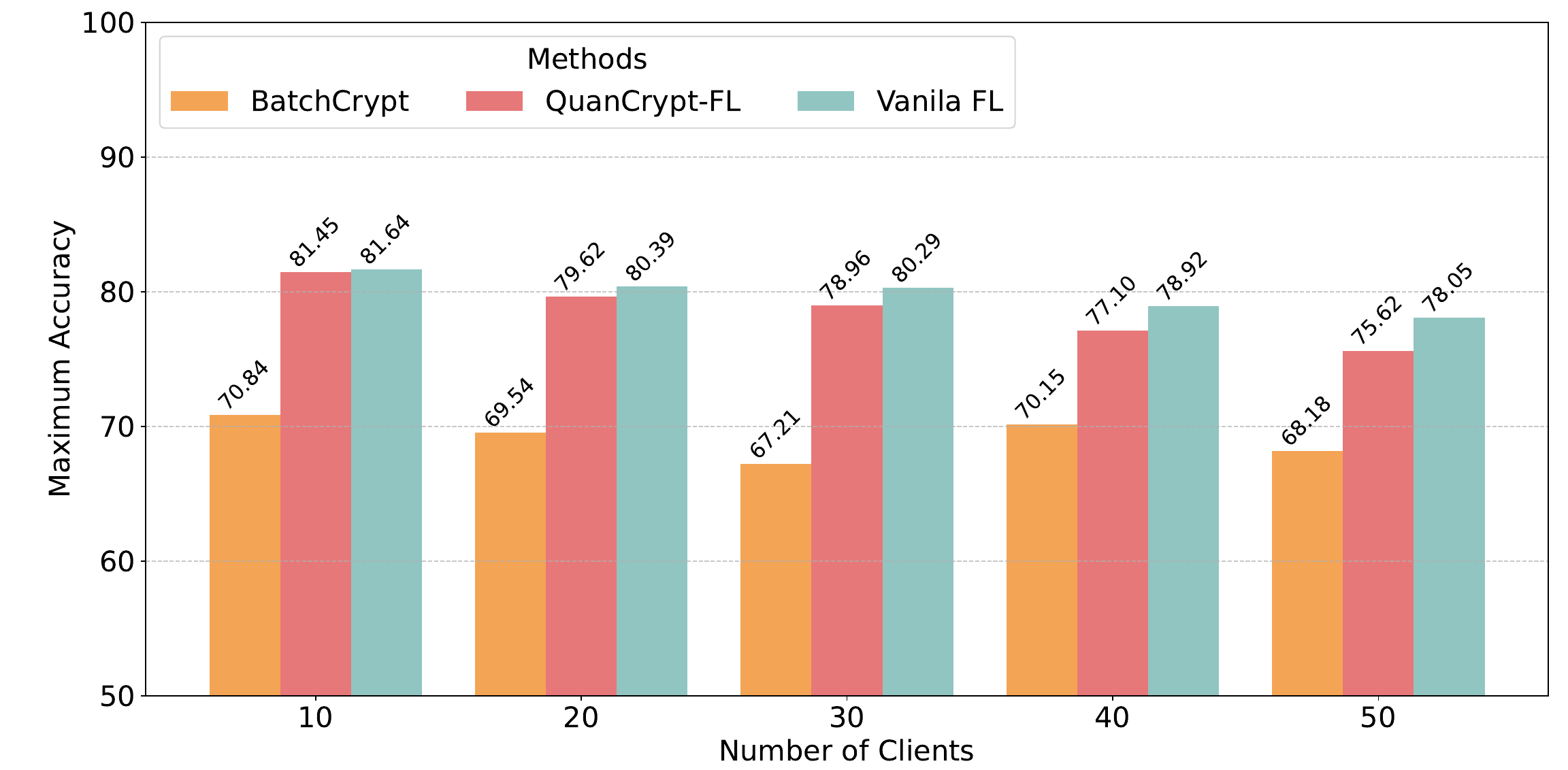}
`    \caption{Comparison of HE mechanism, Model: AlexNet, Dataset: CIFAR10, $\alpha$=3.0, $\lambda$=1.0.}
    \label{fig:c10_bar_com}
\end{figure*}

\begin{table*}[htbp]
\centering
\tiny
\caption{Comparative performance analysis.}
\begin{tabular}{|c|c|c|c|c|}
\hline
\textbf{Method} & \textbf{Model} & \textbf{Dataset} & \textbf{Number of Clients} & \textbf{Maximum Test Acc} \\ \hline

\multirow{5}{*}{\centering BatchCrypt} & \multirow{5}{*}{\centering AlexNet} & \multirow{5}{*}{\centering CIFAR-10} & 10 & 70.84\% \\ \cline{4-5}
 &  &  & 20 & 69.54\% \\ \cline{4-5}
 &  &  & 30 & 67.21\% \\ \cline{4-5}
 &  &  & 40 & 70.15\% \\ \cline{4-5}
 &  &  & 50 & 68.18\% \\ \hline

\multirow{5}{*}{\centering QuanCrypt-FL} & \multirow{5}{*}{\centering AlexNet} & \multirow{5}{*}{\centering CIFAR-10} & 10 & 81.45\% \\ \cline{4-5}
 &  &  & 20 & 79.62\% \\ \cline{4-5}
 &  &  & 30 & 78.96\% \\ \cline{4-5}
 &  &  & 40 & 77.10\% \\ \cline{4-5}
 &  &  & 50 & 75.62\% \\ \hline

\multirow{5}{*}{\centering Vanilla-FL} & \multirow{5}{*}{\centering AlexNet} & \multirow{5}{*}{\centering CIFAR-10} & 10 & 81.64\% \\ \cline{4-5}
 &  &  & 20 & 80.39\% \\ \cline{4-5}
 &  &  & 30 & 80.29\% \\ \cline{4-5}
 &  &  & 40 & 78.92\% \\ \cline{4-5}
 &  &  & 50 & 78.05\% \\ \hline

\multirow{5}{*}{\centering BatchCrypt} & \multirow{5}{*}{\centering AlexNet} & \multirow{5}{*}{\centering CIFAR-100} & 10 & 34.58\% \\ \cline{4-5}
 &  &  & 20 & 33.14\% \\ \cline{4-5}
 &  &  & 30 & 33.56\% \\ \cline{4-5}
 &  &  & 40 & 32.87\% \\ \cline{4-5}
 &  &  & 50 & 33.73\% \\ \hline

\multirow{5}{*}{\centering QuanCrypt-FL} & \multirow{5}{*}{\centering AlexNet} & \multirow{5}{*}{\centering CIFAR-100} & 10 & 47.90\% \\ \cline{4-5}
 &  &  & 20 & 43.70\% \\ \cline{4-5}
 &  &  & 30 & 48.81\% \\ \cline{4-5}
 &  &  & 40 & 48.62\% \\ \cline{4-5}
 &  &  & 50 & 45.40\% \\ \hline

\multirow{5}{*}{\centering Vanilla-FL} & \multirow{5}{*}{\centering AlexNet} & \multirow{5}{*}{\centering CIFAR-100} & 10 & 48.59\% \\ \cline{4-5}
 &  &  & 20 & 49.54\% \\ \cline{4-5}
 &  &  & 30 & 49.14\% \\ \cline{4-5}
 &  &  & 40 & 48.93\% \\ \cline{4-5}
 &  &  & 50 & 49.68\% \\ \hline

\multirow{5}{*}{\centering BatchCrypt} & \multirow{5}{*}{\centering ResNet18} & \multirow{5}{*}{\centering CIFAR-100} & 10 & 34.05\% \\ \cline{4-5}
 &  &  & 20 & 33.14\% \\ \cline{4-5}
 &  &  & 30 & 35.74\% \\ \cline{4-5}
 &  &  & 40 & 36.56\% \\ \cline{4-5}
 &  &  & 50 & 33.73\% \\ \hline

\multirow{5}{*}{\centering QuanCrypt-FL} & \multirow{5}{*}{\centering ResNet18} & \multirow{5}{*}{\centering CIFAR-100} & 10 & 48.21\% \\ \cline{4-5}
 &  &  & 20 & 48.89\% \\ \cline{4-5}
 &  &  & 30 & 48.66\% \\ \cline{4-5}
 &  &  & 40 & 49.26\% \\ \cline{4-5}
 &  &  & 50 & 48.70\% \\ \hline

\multirow{5}{*}{\centering Vanilla-FL} & \multirow{5}{*}{\centering ResNet18} & \multirow{5}{*}{\centering CIFAR-100} & 10 & 52.62\% \\ \cline{4-5}
 &  &  & 20 & 53.48\% \\ \cline{4-5}
 &  &  & 30 & 54.08\% \\ \cline{4-5}
 &  &  & 40 & 53.90\% \\ \cline{4-5}
 &  &  & 50 & 54.20\% \\ \hline

\end{tabular}
\label{tab:QU_max_test_acc}
\end{table*}

Figure \ref{fig:c10_bar_com} compares the performance of BatchCrypt, QuanCrypt-FL, and Vanilla-FL on the CIFAR-10 dataset using the AlexNet model, with the x-axis representing the number of clients (10, 20, 30, 40, 50) and the y-axis showing the maximum accuracy achieved over 300 communication rounds. QuanCrypt-FL consistently achieves accuracy close to Vanilla-FL across all client counts. For example, at 10 clients, QuanCrypt-FL achieves an accuracy of 81.45\%, which is almost identical to Vanilla-FL's 81.64\%. As the number of clients increases, QuanCrypt-FL continues to perform strongly, reaching accuracies of 79.62\%, 78.96\%, 77.10\%, and 75.62\% for 20, 30, 40, and 50 clients, respectively. In contrast, BatchCrypt shows a noticeable drop in performance, with accuracies starting at 70.84\% for 10 clients and steadily declining to 68.18\% at 50 clients. While Vanilla-FL remains the top performer with accuracies of 81.64\%, 80.39\%, 80.29\%, 78.92\%, and 78.05\% as the client count increases, the minimal difference between QuanCrypt-FL and Vanilla-FL indicates that QuanCrypt-FL closely approximates Vanilla-FL’s performance while consistently outperforming BatchCrypt across all client counts.

Due to space limitations, we do not provide the result visualization of the CIFAR-100 dataset. Table \ref{tab:QU_max_test_acc} presents the comparative analysis results for all methods across different datasets and models, including CIFAR-10 using AlexNet and CIFAR-100 using AlexNet and ResNet18. The results demonstrate that QuanCrypt-FL outperforms BatchCrypt. For CIFAR-10, we achieve accuracy similar to Vanilla FL, with less than a 1\% accuracy loss, while BatchCrypt shows a greater than 10\% accuracy loss compared to Vanilla FL across different client counts (10, 20, 30, 40, 50). For the CIFAR-100 dataset with the AlexNet model, QuanCrypt-FL achieves accuracy close to Vanilla FL, except for an approximate 5\% loss for 20 clients, whereas BatchCrypt exhibits a greater than 15\% accuracy loss. For the ResNet18 model on CIFAR-100, Vanilla FL achieves around 54\% accuracy, while QuanCrypt-FL reaches approximately 50\%, and BatchCrypt experiences a 20\% accuracy loss compared to Vanilla FL.

\begin{figure*} [!t]
    \centering
    \includegraphics[width=0.9 \linewidth]{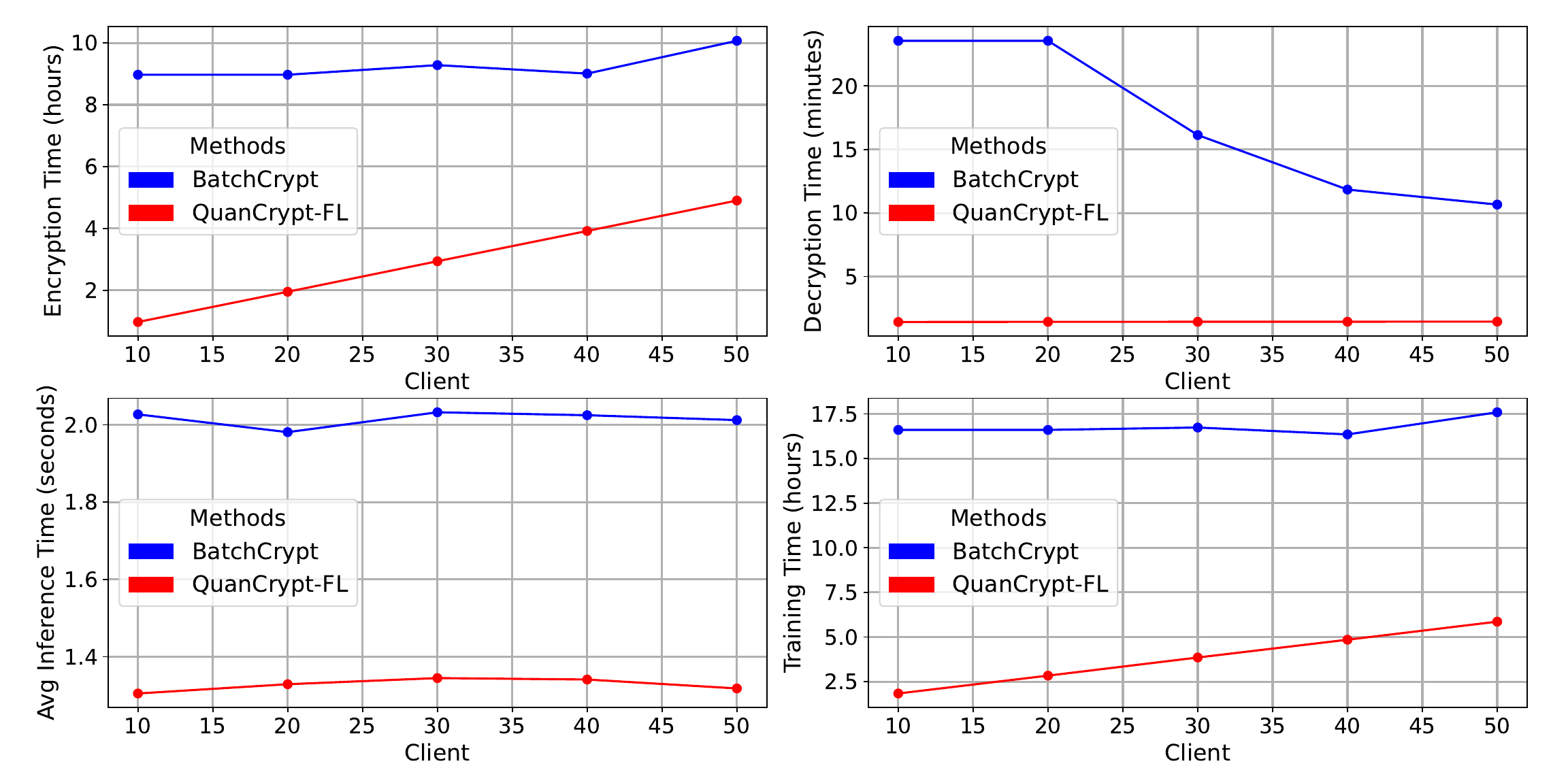}
`    \caption{Comparison of time of HE mechanism, Model: AlexNet, Dataset: CIFAR10, $\alpha$=3.0, $\lambda$=1.0.}
    \label{fig:c10_all_time}
\end{figure*}

\begin{figure}[!t]
    \centering
    \includegraphics[width=0.94\linewidth]{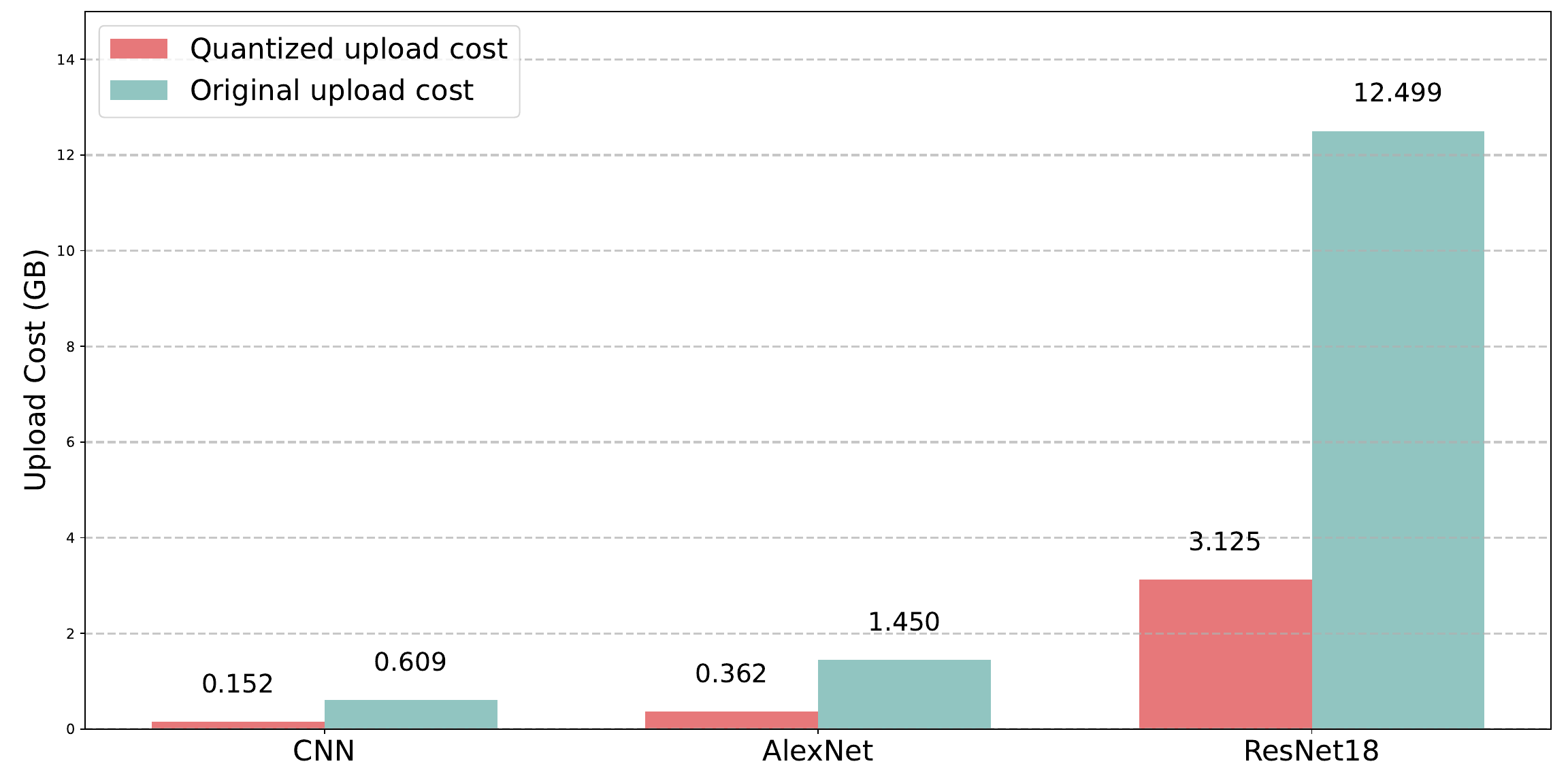}
    \caption{Comparison of upload cost for different models, Dataset: CIFAR10, $\alpha$=3.0, $\lambda$=1.0.}
    \label{u_cost}
\end{figure}

\begin{figure}[htbp]
    \centering
    \begin{subfigure}[b]{0.15\textwidth}  
        \centering        \includegraphics[width=\textwidth]{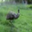}
        \caption{}
    \end{subfigure}
    \begin{subfigure}[b]{0.15\textwidth}  
        \centering
        \includegraphics[width=\textwidth]{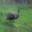}
        \caption{}
    \end{subfigure}
     \begin{subfigure}[b]{0.15\textwidth}
        \centering
        \includegraphics[width=\textwidth]{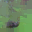}
        \caption{}
    \end{subfigure}
    
    \caption{Comparison of image reconstruction using GIA, Model: ResNet18, Dataset: CIFAR10, $\alpha$=3.0, $\lambda$=1.0. (a) Original image, (b) Gradient inversion attack on the global model in Vanilla FL, (c) Gradient inversion attack on global model in QuanCrypt-FL}
    \label{attack_simulation}
\end{figure}

Figure \ref{fig:c10_all_time} presents the comparison between BatchCrypt and QuanCrypt-FL in terms of encryption time, decryption time, average inference time, and training time, as the number of clients increases (10, 20, 30, 40, 50). The results reveal significant differences in performance with the existing methods, with QuanCrypt-FL demonstrating considerable time efficiency across all metrics.

For \textbf{encryption time}, BatchCrypt consistently requires more time as the number of clients increases, ranging from 8.97 hours at 10 clients to 10.07 hours at 50 clients. In contrast, QuanCrypt-FL shows much shorter encryption times, ranging from 0.97 hours at 10 clients to 4.90 hours at 50 clients. This indicates that QuanCrypt-FL is \textbf{up to 9 times faster} than BatchCrypt in encryption time, particularly as the number of clients grows.

In terms of \textbf{decryption time}, BatchCrypt decreases its time from 23.55 minutes at 10 clients to 10.66 minutes at 50 clients. However, QuanCrypt-FL maintains a much more efficient and stable decryption time, ranging from 1.42 minutes at 10 clients to 1.45 minutes at 50 clients, making it \textbf{up to 16 times faster} than BatchCrypt.

When comparing \textbf{average inference time}, BatchCrypt stays relatively stable, with values around 2.02 seconds, whereas QuanCrypt-FL achieves faster inference times, ranging from 1.30 to 1.34 seconds. This demonstrates that QuanCrypt-FL is \textbf{about 1.5 times faster} in inference compared to BatchCrypt.

Finally, for \textbf{training time}, BatchCrypt requires between 16.60 and 17.59 hours as the client count increases, while QuanCrypt-FL significantly reduces training time, from 1.83 hours at 10 clients to 5.86 hours at 50 clients. This makes QuanCrypt-FL \textbf{up to 3 times faster} in training compared to BatchCrypt.

In terms of storage, we calculate communication costs for model uploads as shown in Figure \ref{u_cost}. During uploads, we apply quantization to compress model parameters into an 8-bit format, significantly reducing upload costs. For downloads, clients typically receive the full model in a 32-bit format, which increases communication costs. To address this, pruning can be considered an effective approach to reduce costs. Although our mechanism currently uses a 32-bit format for download costs, pruning can reduce costs by setting less important weights to zero after aggregation and encoding sparse parameters in a 4-bit format, while retaining non-sparse parameters in 32-bit format. While we applied pruning, we did not visualize the download cost in our results. This approach effectively reduces both upload and potential download costs while preserving model performance.

We simulated a Gradient Inversion Attack (GIA) on both our proposed QuanCrypt-FL model and the Vanilla FL model. For the attack simulation, we used 10 clients and a ResNet18 model and CIFAR10 dataset in both methods. Apart from enhancing efficiency, pruning is also employed as a defense mechanism against attacks in our approach. By reducing the model's complexity through pruning, we introduce sparsity, which limits the success of attacks like GIA. This dual benefit of pruning not only improves computational performance but also strengthens the model’s resistance to adversarial attacks, ensuring greater security during the training process.
Specifically, we set the initial pruning rate at 30\% and progressively increased it to a target pruning rate of 70\%. 

In Figure \ref{attack_simulation}, subfigures (a) original images, while subfigure (b) shows the reconstructed image generated by the GIA attack in Vanilla FL, as described in the threat model section. In Vanilla FL, the attack was nearly 100\% successful in reconstructing the original image from the gradients. However, in our QuanCrypt-FL method, due to the sparsity introduced by the pruning technique, the GIA was unsuccessful in reconstructing the original image from the gradients in figure (c). 

This demonstrates that, even if the server or adversary has access to the shared model parameters, it cannot infer sensitive information or reconstruct data from the shared gradients or even from the decrypted global model used for inference and evaluation.

%% file: Discussion.tex
\section{Discussion} \label{sec:Discussion}
Our proposed mechanism, QuanCrypt-FL, achieves state-of-the-art performance compared to existing methods that implement HE with pruning and quanization in FL. Compared to BatchCrypt’s approach of batching quantized gradient values into smaller units, our layer-wise encryption strategy in FL significantly reduces the frequency of encryption and decryption operations required per training round. By encrypting entire layers instead of smaller gradient batches, we streamline the encryption process, resulting in faster overall FL training times. This approach also reduces computational overhead during aggregation, as each layer can be processed as a single encrypted unit, simplifying the update process. Consequently, our method enhances the efficiency and scalability of secure FL, making it a practical choice for applications prioritizing both speed and data privacy. We present a detailed comparative analysis with the BatchCrypt. While BatchCrypt can achieve similar accuracy to our method on simpler datasets like MNIST, it shows significant limitations with larger and more complex datasets such as CIFAR-100. For example, in our experiments using CIFAR-100 with the AlexNet model, BatchCrypt achieved very low accuracy, whereas our method reached accuracy levels close to Vanilla FL. In this setup, we used an IID data distribution and allocated 5000 data samples per client.

Moreover, we observed that BatchCrypt incurs considerably higher computational complexity compared to QuanCrypt-FL. We achieved faster encryption, decryption, inference, and training time. Moreover. our method requires less memory, reducing the upload cost for clients during FL training. Additionally, we implemented iterative pruning to create a progressively sparse model during the communication rounds, which reduces inference time. As a result, QuanCrypt-FL demonstrates faster inference speeds than both BatchCrypt and Vanilla FL. Furthermore, pruning also strengthens defenses against GIA; a semi-honest server or adversary cannot infer sensitive information, even with access to client model updates or the global model. If we increase the pruning rate beyond 70\%, our defense technique becomes more resilient against GIA attacks under the adversarial scenario or in the case of an honest but curious server. However, we need to maintain a tradeoff between the pruning rate and utility to make our mechanism feasible for real-world scenarios.